\documentclass[runningheads]{llncs}
\usepackage{graphicx}

\usepackage{tikz}
\usepackage{comment}
\usepackage{amsmath,amssymb} 
\usepackage{color}

\usepackage[accsupp]{axessibility}  

\begin{document}
\pagestyle{headings}
\mainmatter
\def\ECCVSubNumber{2891}  

\title{Seeing Far in the Dark with Patterned Flash} 

\titlerunning{Seeing Far in the Dark with Patterned Flash}
%

\author{Zhanghao Sun\inst{1}$^{*\dagger}$ \and
Jian Wang\inst{2}$^*$ \and
Yicheng Wu\inst{2} \and
Shree Nayar\inst{2}}
\authorrunning{Zhanghao Sun, Jian Wang, Yicheng Wu, Shree Nayar}
%
\institute{Stanford University, 350 Serra Mall, Stanford, CA 94305, USA \and
Snap Inc., 229 W 43rd St, New York, NY 10036, USA}
\maketitle

\def\thefootnote{*}\footnotetext{These authors contributed equally.}\def\thefootnote{\arabic{footnote}}
\def\thefootnote{$\dagger$}\footnotetext{This work was done during an internship at Snap Research.}\def\thefootnote{\arabic{footnote}}

\begin{abstract}
Flash illumination is widely used in imaging under low-light environments. However, illumination intensity falls off with propagation distance quadratically, which poses significant challenges for flash imaging at a long distance. We propose a new flash technique, named ``patterned flash'', for flash imaging at a long distance. Patterned flash concentrates optical power into a dot array. Compared with the conventional uniform flash where the signal is overwhelmed by the noise everywhere, patterned flash provides stronger signals at sparsely distributed points across the field of view to ensure the signals at those points stand out from the sensor noise. This enables post-processing to resolve important objects and details. Additionally, the patterned flash projects texture onto the scene, which can be treated as a structured light system for depth perception. Given the novel system, we develop a joint image reconstruction and depth estimation algorithm with a convolutional neural network. We build a hardware prototype and test the proposed flash technique on various scenes. The experimental results demonstrate that our patterned flash has significantly better performance at long distances in low-light environments. Our code and data are publicly available.\footnote{https://github.com/zhsun0357/Seeing-Far-in-the-Dark-with-Patterned-Flash} 


\end{abstract}

\begin{figure}[t!]
\begin{center}
\includegraphics[width=0.9\linewidth]{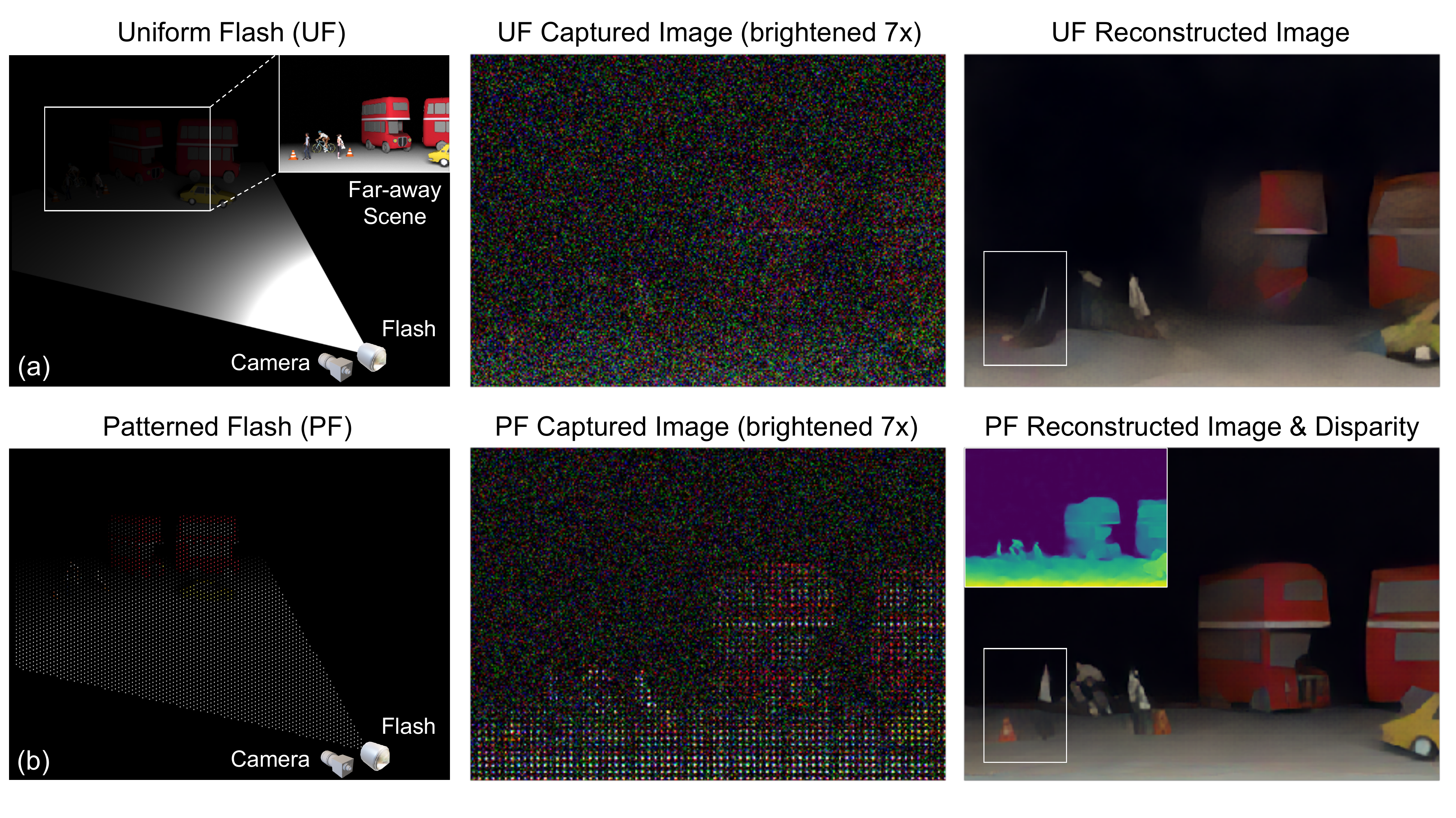}
\end{center}
   \caption{(Better viewed on full screen to avoid visual artifacts) Conventional, uniform flash (UF) vs. the proposed patterned flash (PF). (a) In UF system, due to inverse square law of light, only close-by scenes are lit up, while the far-away scene is overwhelmed by sensor noise. The ground truth scene is in the inset of the left column. (b) In PF system, illumination power is concentrated into a dot array to enable long distance imaging in low-light environments. The reconstructed image contains important objects and details that are missing in UF reconstructed image (e.g., the pedestrian and the traffic cone in the white box). PF also supports depth estimation with the textured illumination pattern, as shown in the inset of the right column.}
\label{teaser}
\end{figure}

\section{Introduction}
\label{intro}
Low-light imaging is critical in consumer photography, surveillance, robotics, and vision-based autonomous driving. Both hardware designs~\cite{back_illum},  \cite{yang2004high_microlens}, \cite{ryyb}, \cite{hubel2004spatial_specialsensor}, \cite{flint2017flash}, \cite{nir_flash}, \cite{jung2017enhancement_rgb_mono}, \cite{quanta_burst} and computational imaging algorithms \cite{dabov2007image_BM3D}, \cite{sid}, \cite{brooks2019unprocessing}, \cite{google_hdr+}, \cite{pixel_hdr} have been developed for this challenging task. Among these approaches, flash illumination is widely adopted given its outstanding performance and robustness. A flash can easily boost the illumination intensity hence the image signal level by $100 \sim 1000\times$ \cite{flint2017flash}, \cite{altenburg2012understanding}. With high enough illumination level, it is sufficient to use short exposure time, casual capturing mode, and simple image processing while still getting high-quality images. However,  flash has a fundamental drawback; its strength attenuates quadratically with distance. With a limited power budget (which is the practical constraint for most mobile devices like phones, drones, and wearable devices), the conventional flash illumination system is ineffective at a long distance. This raises robustness and safety issues in surveillance and navigation (e.g., missing important objects at night time). It also leads to irreversible content degradation in photography.


In this paper, we propose patterned flash (PF) technique for low-light imaging at a long distance. Instead of distributing the flash light uniformly across the camera's field of view (FOV), we concentrate it into a dot array pattern. Considering the same power budget, in a conventional uniform flash (UF) system, flash signals from far-away objects are overwhelmed by the sensor noise or fall below the sensitivity threshold of the sensor, leading to severe information loss (as shown in Fig.~\ref{teaser}(a)). Contrarily, in the PF system (Fig.~\ref{teaser}(b)), the perceived dot array contains higher signals from far-away objects, despite being sparse. By leveraging the redundancies in natural images and capabilities of modern neural networks, we can reconstruct far-away scenes with higher quality. As shown in the third column of Fig.~\ref{teaser}, PF's reconstruction of red buses is better; more importantly, the pedestrian and the traffic cone (indicated by the white box) are completely missed in UF's result while still visible in PF's. Downstream tasks, including object detection~\cite{centernet}, \cite{obj_det_lowlight} and semantic segmentation~\cite{sem_seg_denoise}, can also benefit from this higher quality reconstruction. Note that concentrating optical power has also been proposed to improve the sensing distance of active 3D sensors~\cite{episcan}, \cite{epitof}, 
\cite{wang2018programmable}, \cite{sunlight_si}, but in that context, photon noise from strong ambient light sources (e.g., sunlight), instead of sensor noise, dominates.

Apart from higher image quality at a long distance, PF introduces another advantage over UF. With a textured illumination pattern and a small baseline between the light source and the imaging device, the captured image contains depth information, as in the case of structured light (SL) 3D imaging techniques \cite{geng2011structured}. The slight differences to traditional SL are that our SL's pattern is weak in brightness, and ours has a comparatively micro baseline because of the long imaging distance. Since color and depth information highly correlate in the captured image, existing SL methods cannot be readily applied. The proposed algorithm solves this problem and can estimate sub-pixel disparity accurately. 

\paragraph{\indent{\normalfont In summary, our contributions are in three folds:}}
\begin{itemize}
    \item We propose patterned flash (PF) for low-light imaging. It relieves flash's fundamental drawback, short distance.
    \item We develop a deep-learning-based joint image reconstruction and disparity estimation algorithm to reconstruct scene image and depth from the single captured patterned flash image.
    \item  We evaluate our approach with simulations and real-world data captured by a hardware prototype. The results verify that the proposed system has significant improvement over the conventional uniform flash.
\end{itemize}

\section{Related Work}
\label{related}
\subsection{Flash Imaging}
The flash has long been used to enhance image signals in dim environments. It was also applied in many other computational imaging tasks, such as reflection removal \cite{lei2021robust}, BRDF acquisition \cite{hui2017reflectance}, white balance \cite{hui2016white}, illumination separation \cite{hui2018illuminant}, image matting~\cite{flash_mat}, edge detection \cite{raskar2004non} and
geometric estimation \cite{li2018learning}. In the context of using a flash to assist low-light imaging, there are three well-known problems, and some solutions have been proposed. (1) Flash images look harsh or unnatural. Flash changes the color tone of the scene, adds highlights to the human face, and makes the nearby objects much brighter than far away. Since the seminal work~\cite{fnf1}, \cite{fnf2}, flash/no-flash techniques have been intensively discussed in literature \cite{fnf3}, \cite{fnf4}, \cite{guo2018mutually}, \cite{zhihao} which denoise the no-flash image with guidance from the flash image. True-tone flash has been widely used in smartphones which uses two LEDs of different colors to mimic the color tone of the scene.
(2) Flash is intrusive to human eyes. 
To make the flash more user-friendly and social friendly, previous research focused on using flashes that operate at wavelengths less sensitive to human eyes, e.g.,  dark flash (NUV and NIR) \cite{nir_flash}, \cite{nir_flash1}, deep-red flash \cite{jinhui}, or using weak flash \cite{Huawei_lowflash}. (3) Flash has a short effective range. Current methods either increase the power or shrink the  FOV, and there has been little effort to solve this problem from a computational imaging perspective. High power (10 $\sim$ 100 Watts) LEDs have been used to lighten large scenes at night (e.g., football field), but such systems cannot be applied to consumer devices and introduce heavy light pollution. Another approach is to concentrate light into a smaller FOV. Matrix headlight has been adopted in high-end vehicle models~\cite{matrix_headlight}; when necessary, it concentrates flash power into a smaller angular range to reach a longer distance. However, sacrificing FOV is a non-optimal solution.

More broadly, flash illumination or active illumination is also often used in other imaging modalities, where the environmental light level is low, including active hyper-spectral, short-wave infrared \cite{steiner2016design}, middle-wave infrared \cite{koerperick2009high}, long-wave infrared \cite{koerperick2010cascaded}, and near-ultraviolet imaging. These techniques also face the challenge of light fall off, and they may face more challenges than active, visible-light imaging in that the sensor qualities \cite{rogalski2019infrared} and light emitters' optical efficiencies \cite{tan2011optical} are lower, or the power is constrained by safety concerns.

%

\subsection{Low-light Imaging without Flash}
Apart from flash imaging, there are other popular low-light imaging methods, which can be categorized into single image-based and burst-based methods. Usually, they require more ambient light than the flash-based methods. Efforts have been made to denoise a single image from early years of image processing to date \cite{dabov2007image_BM3D}, \cite{sid}, \cite{brooks2019unprocessing}, \cite{physical_noise}. Chen et al.~\cite{sid} first introduced a modern neural network model on RAW data for low-light imaging. Wei et al. ~\cite{physical_noise}  demonstrated that carefully modeling sensor noise in the training data plays an important role in improving the denoising performance. Though progress has been made, it is fundamentally difficult to recover the scene details without hallucination in single-image-based methods. Burst-based imaging has achieved more impressive results recently \cite{google_hdr+}, \cite{pixel_hdr}, \cite{burst1}, \cite{burst2}. By fusing a sequence of images taken with a short exposure time, SNR is greatly boosted. The major difficulty for this approach is aligning the burst frames; some algorithms are based on explicit optical flow estimation~\cite{google_hdr+}, \cite{pixel_hdr} while others utilize implicit kernel prediction networks~\cite{burst1}, \cite{burst2}; for both approaches, the extremely low-light condition makes the image registrations unreliable, and then the reconstruction quality reduces significantly~\cite{burst1}. Burst-based methods also require the camera to be still and the scene to be static, which may not hold in some circumstances.

\subsection{Structured Light (SL) 3D Imaging}
SL is one of the most widely used techniques in 3D imaging \cite{geng2011structured}. Although there are advances which can speed up dot scanning \cite{wang2016dual}, line scanning \cite{bartels2019agile}, and time-multiplexing gray code \cite{sun2022structured}, the random dot pattern is usually adopted in consumer devices, like Microsoft Kinect \cite{maccormick2011does} and Apple Face ID \cite{mainenti2017user}, because it supports single-shot depth estimation. Existing algorithms, like block matching or HyperDepth \cite{fanello2016hyperdepth}, assume the scene of interest is close, the baseline is sufficiently large, and the dots are saturated in the image. Depth can be accurately estimated by the number of pixel shifts of pattern dots. However, these algorithms can not be directly applied in our small-baseline system, since the sub-pixel disparities mainly lead to subtle color and intensity changes. Thus, the depth and image estimations are highly correlated. Saragadam et al.~\cite{mbsl} proposed micro-baseline SL to estimate sub-pixel disparities, but it requires a triangular wave pattern profile. Riegler et al.~\cite{connect_dots} proposed a self-supervised learning approach by modeling the joint distribution of scene albedo and depth, but it requires sufficient ambient illumination to separate scene albedo and depth. Low-light environments make this separation impractical. The proposed PF is a novel SL system where color and depth information are naturally mixed. Estimating either attribute requires knowledge about the other one, and the baseline is very small. We propose a new algorithm that can estimate both accurately. 


\section{Image Formation Model}
In the proposed PF system, we capture one RAW image $\mathbf{I} \in \mathbb{R}^{H\times W\times C}$, with the PF as the dominant illumination light source\footnote{Our method can be extended to the scenarios with stronger ambient illumination. Please refer to supplementary material for details.} (with $H$, $W$ being spacial dimensions and $C$ being the number of color channels). A reference pattern $\mathbf{P_r} \in \mathbb{R}^{H\times W\times C}$ is captured in the  calibration stage, and the image/reference pair $\{\mathbf{I}$, $\mathbf{P_r}\}$ is used as the  input into the reconstruction algorithm. Both $\mathbf{I}$ and $\mathbf{P_r}$ are $\in [0,1]$. The image formation model for $\mathbf{I}$ can be expressed as:
\begin{equation}
\label{eqn:image_form}
    \mathbf{I} = \frac{\mathcal{W}_{\mathbf{d}}[\mathbf{P_r}]}{\mathbf{d}^2} \odot \mathbf{A} + \mathcal{N}
\end{equation}


In Eq.~\ref{eqn:image_form}, $\mathcal{W}_\mathbf{d}$ is the depth-dependent warping operation. $\mathcal{W}_\mathbf{d}[\mathbf{P_r}]$ is the warped flash pattern. $\mathbf{d}$ is the scene depth, and $\mathbf{A}$ is the scene image. Here we include scene surface normal and material properties in $\mathbf{A}$. $\odot$ is the pixel-wise product. $\mathcal{N}$ includes photon noise, read noise, quantization noise, and other noise sources~\cite{physical_noise}. We leverage this image formation model in synthetic data generation, for network training and evaluation. We also implement a photometric loss based on this model for better supervision.

\begin{figure}[t!]
\begin{center}
\includegraphics[width=0.9\linewidth]{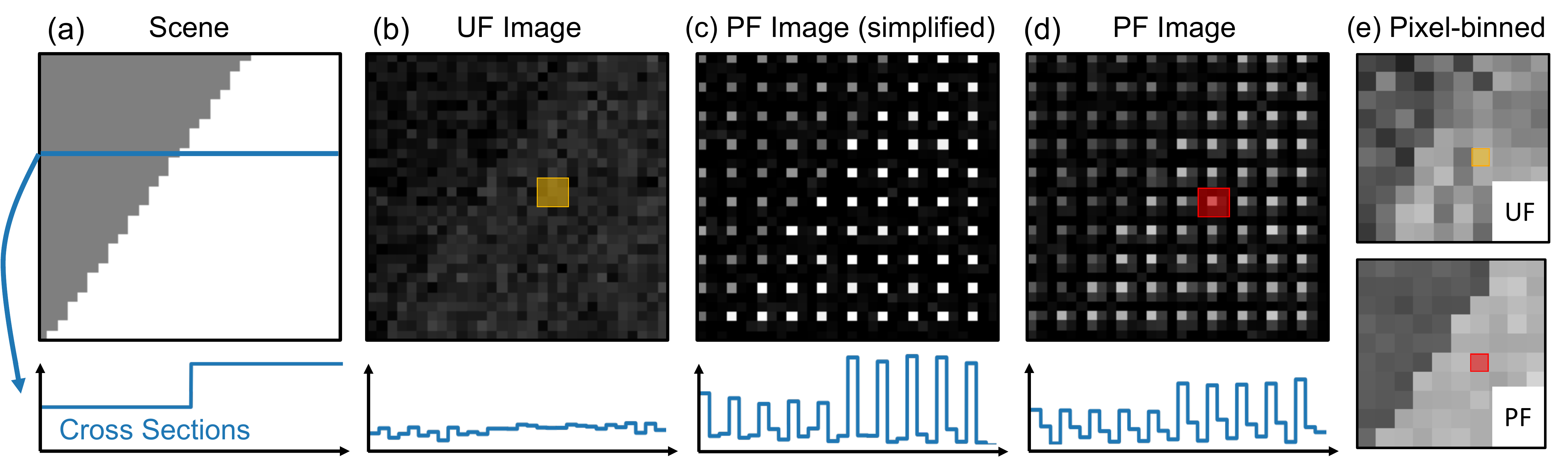}
\end{center}
   \caption{Signal-noise-ratio comparisons and ``pixel-binned'' images. (a) Scene. (b-d) Comparisons between UF and PF captured images with cross-sections. (e) Comparisons between ``pixel-binned'' UF and PF images, as discussed in Sec. ~\ref{sec:theory}. The yellow and red pixels in upper and lower parts of (e) correspond to the $M \times M$ patches ($M = 3$ in this case) in (b) and (d), respectively.}
\label{imf}
\end{figure}

\subsection{Signal-to-noise Ratio Analysis}
\label{sec:theory}
Here we theoretically analyze the advantage of PF over UF and elucidate the assumptions used in the comparison. In the UF system (Fig.~\ref{imf}(b)), each pixel receives signal $s$, the signal-noise-ratio at each pixel can be expressed as:
\begin{equation}
\label{eqn:snru}
    SNR_{UF} = s/\sqrt{s + \sigma_r^2}
\end{equation}
where $\sigma_r$ is the standard deviation of signal-independent sensor noise (e.g., read noise), and photon shot noise of signal $s$ has variance $s$. 

In the PF system, however, suppose the entire signal in an $M\times M$ patch is concentrated in one camera pixel (Fig.~\ref{imf}(c), $M = 3$). The ``in-pattern'' pixel receives signal $M^2s$, and the SNR for this pixel can be expressed by $SNR_{PF}$ in Eq.~\ref{eqn:snrp}. However, this SNR enhancement is at the expense of lower spatial resolution. For a fair comparison, we average the entire signal in the UF image patch to get the SNR at the same spatial resolution ($SNR_{UFavg}$ in Eq.~\ref{eqn:snrp}). We notice that when $\sigma_r$ dominates, $SNR_{PF}$ can be around $M$ times higher than $SNR_{UFavg}$. Generally, this holds in low-light conditions~\cite{noise1}, \cite{noise2}.

\begin{eqnarray}
\label{eqn:snrp}
\left\{ \begin{array}{l}
    SNR_{PF} = M^2s/\sqrt{M^2s + \sigma_r^2} \approx M^2s/\sigma_r \\
    SNR_{UFavg} = M^2s/\sqrt{M^2s + M^2\sigma_r^2} \approx Ms/\sigma_r
\end{array} 
\right.
\end{eqnarray}

For real-world patterns, $\mathbf{P_r}$ may not be exactly ``$1$ in $M\times M$'' (Fig.~\ref{imf}(d)). Each pattern dot is blurred, and the overall pattern contrast is lower. We can define a similar input signal SNR comparison. For a PF image, we set a threshold to define whether an image pixel falls within a pattern dot. Then we only extract ``in pattern'' image pixels from each $M\times M$ pixels patch (red rectangle in Fig.~\ref{imf} (d)) and combine them into a low-resolution image (as shown in Fig.~\ref{imf}(e), lower part). For a UF image, we average the captured image to the same spatial resolution (as shown in Fig.~\ref{imf}(e), upper part). In the following, we use this ``pixel-binning'' method to visualize the quality of image signal before processing. 

We also define the ``average occupancy'' of $\mathbf{P_r}$ as its mean value. A pattern with average occupancy $1/\Omega$ can be regarded as an equivalence of the simplified ``$1$ in $\sqrt{\Omega} \times \sqrt{\Omega}$'' pattern, and therefore has theoretical upper bound of input SNR gain $\sqrt{\Omega}\times$ (when signal $\to 0$ and contrast $\to \infty$). With smaller average occupancy, the pattern is sparser, and the SNR gain is larger. However, recovering the image content between pattern dots becomes more difficult. Empirically, we found average occupancy $=1/16$ achieves good balance in this trade-off. Note that higher photon shot noise and lower flash pattern contrast would reduce the SNR gain from its theoretical limit. However, we demonstrate that, in both simulations and real-world experiments,  PF improves image quality significantly. 

\section{Patterned Flash Processing}

\begin{figure}[t!]
\begin{center}
\includegraphics[width=0.9\linewidth]{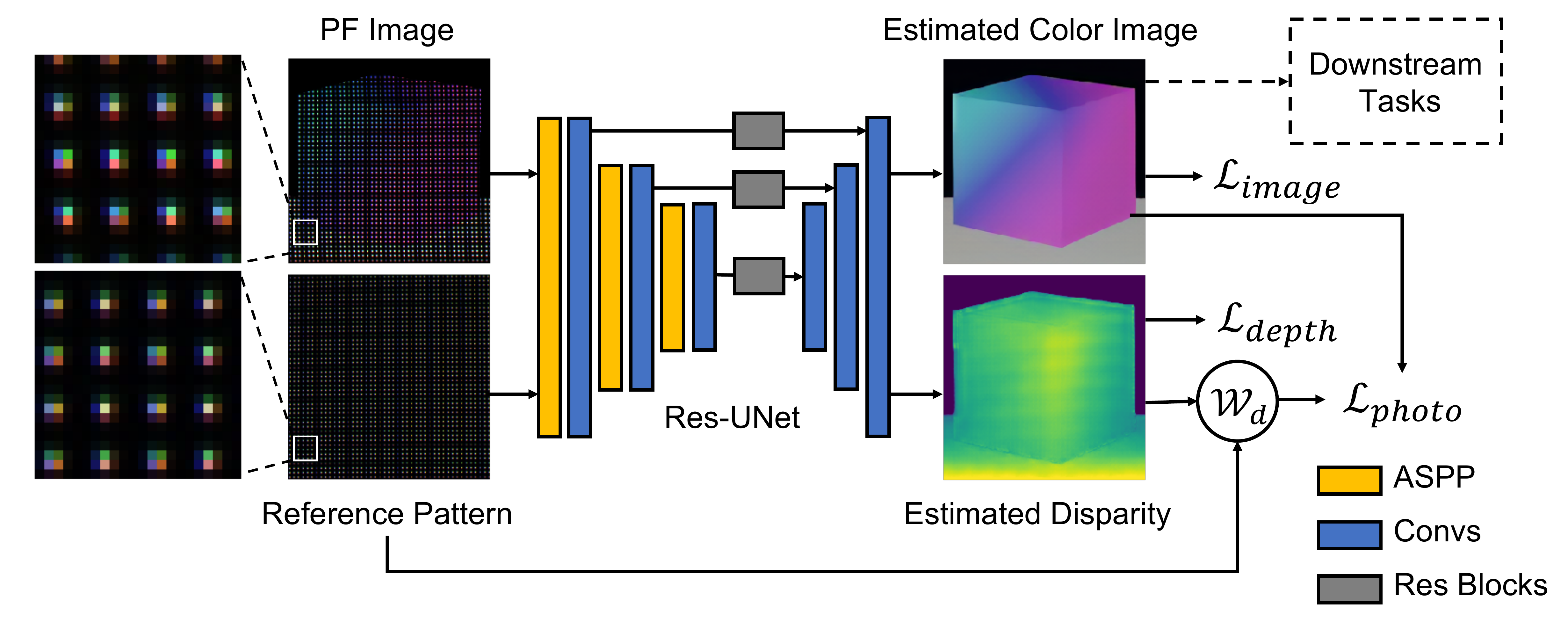}
\end{center}
   \caption{Joint scene image and depth estimation. We use captured PF image and a reference pattern as input. The neural network predicts a scene image and a disparity map. We use three loss functions to supervise the network training: image loss $\mathcal{L}_{image}$, depth loss $\mathcal{L}_{depth}$, and photometric loss $\mathcal{L}_{photo}$. The reconstruction results can be further employed in downstream tasks (e.g., object detection).}
\label{pipeline}
\end{figure}

\subsection{Network Architecture}
The network architecture is shown in Fig.~\ref{pipeline}. The goal of our reconstruction algorithm is to jointly estimate a scene image $\widehat{\mathbf{A}}$ and the scene depth $\widehat{\mathbf{d}}$ from PF image $\mathbf{I}$. We use a U-Net~\cite{unet} as backbone, with multiple residual blocks~\cite{resnet} as skip connections. Since information is distributed sparsely in $\mathbf{I}$, a big receptive field is required for feature extraction. We use an Atrous Spatial Pyramid Pooling (ASPP) module~\cite{aspp} to balance the receptive field and computational cost. For details of the network architecture, please refer to supplementary material.

\subsection{Loss Functions}
We train the network in a supervised manner, with ground truth color image $\mathbf{A}$ and depth $\mathbf{d}$. We use three types of loss functions; supervision on image reconstruction $\mathcal{L}_{image}$, supervision on depth estimation $\mathcal{L}_{depth}$, and photometric loss~\cite{connect_dots}, \cite{si1} $\mathcal{L}_{photo}$. The total loss is a weighted combination.
\begin{eqnarray}
\mathcal{L} = \mathcal{L}_{image} + \lambda_1 \mathcal{L}_{depth} + \lambda_2 \mathcal{L}_{photo}
\end{eqnarray}
The image reconstruction loss consists of three parts, an L2 loss, a perceptual loss~\cite{percep_loss} (with VGG16 as feature extractor), and an L1 gradient loss.
\begin{eqnarray}
\mathcal{L}_{image} = |\widehat{\mathbf{A}} - \mathbf{A}|^2 + \beta_1 |\textrm{VGG}(\widehat{\mathbf{A}}) - \textrm{VGG}(\mathbf{A})|^2 + 
\nonumber\\
\beta_2 (|\nabla_x\widehat{\mathbf{A}} - \nabla_x \mathbf{A}| + |\nabla_y \widehat{\mathbf{A}} - \nabla_y \mathbf{A}|)
\end{eqnarray}
For depth estimation supervision, we use an L2 loss.
\begin{eqnarray}
\mathcal{L}_{depth} = |\widehat{\mathbf{d}} - \mathbf{d}|^2
\end{eqnarray}
The photometric loss utilizes the image formation model to enforce a consistent depth and image prediction.
\begin{eqnarray}
\mathcal{L}_{photo} = |\mathcal{W}_{\widehat{\mathbf{d}}}[\mathbf{P_r}] \odot \widehat{\mathbf{A}} - 
\mathcal{W}_{\mathbf{d}}[\mathbf{P_r}] \odot \mathbf{A}|^2
\end{eqnarray}

\section{Implementations}
We train the network on data synthesized from FlyingThings3D dataset~\cite{ft3d}, with ground truth image and depth. We use experimentally captured reference patterns for both network training and evaluations. For each scene, we set a distance $\in [8, 16]$. Compared with imaging an object at distance $=1$, this corresponds to an attenuation of the color image (normalized to $[0,1]$) by $64\times \sim 256\times$. 
Note that we use a uniform attenuation factor across one scene in training data synthesis instead of a depth-dependent attenuation $1/\mathbf{d}^2$. This is to not weigh the farther objects less in loss function $\mathcal{L}_{image}$. We assume a heteroscedastic Gaussian noise model to account for both read and shot sensor noise. 
For an attenuated image $\mathbf{I_0}$, the noise variance is defined as $\sigma_r^2 + \sigma_s^2\mathbf{I_0}$. For each scene, we randomly select $\sigma_r \in [0.002, 0.005]$ and $\sigma_s \in [0.015, 0.04]$. 
We also add a small ``row noise'', with $\sigma_{row} = 0.0005$, to be consistent with our CMOS sensor used in real-world experiments~\cite{physical_noise}. For each scene, we randomly choose the sign of disparities and randomly select a baseline between $0.5 \sim 5.0$, consistent with the unambiguous range defined by the reference pattern.

We train our model using the Adam optimizer~\cite{adam}, with learning rate $2\times10^{-4}$, for a total of roughly $200$k iterations. The training process takes around 1.5 days on a single Nvidia Tesla-V100 GPU.

\begin{figure}[t!]
\begin{center}
\includegraphics[width=1.0\linewidth]{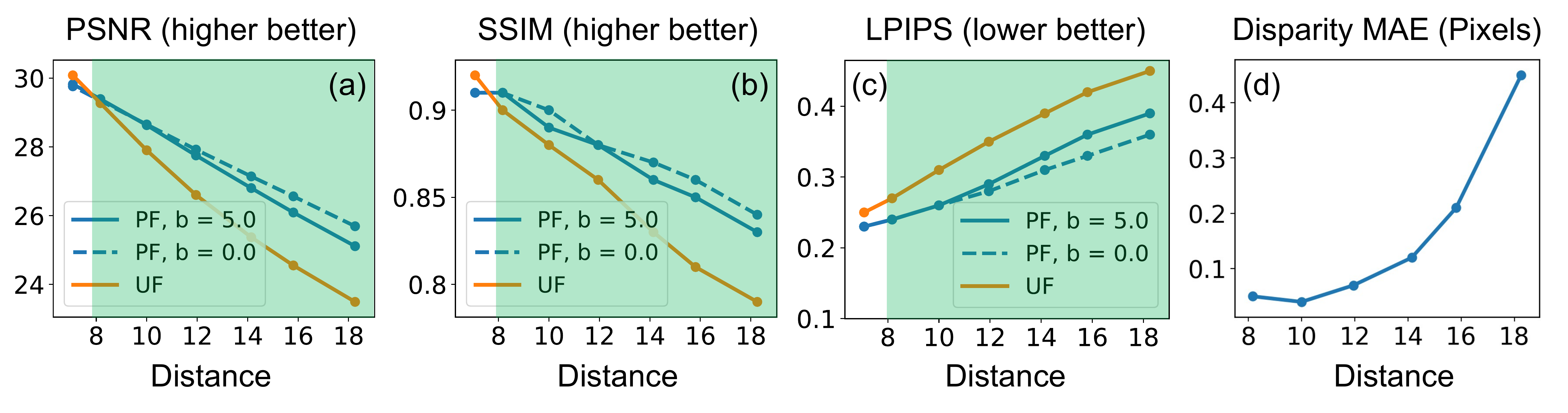}
\end{center}
   \caption{Quantitative comparisons between PF and UF on synthetic data (distance unit defined in Sec.~\ref{sec:simu_result_sub1}). (a, b) Image reconstruction quality from the PF system consistently outperforms that in the UF system at a long distance (green region). Also, increasing the baseline only slightly influences the image reconstruction performance. (c) Mean absolute error (MAE) of disparity estimation.}
\label{quantitative}
\end{figure}

\section{Simulation Results}
\label{sec:simu_result}
\subsection{Joint Image and Depth Estimation}
\label{sec:simu_result_sub1}
We first evaluate the image reconstruction quality at discrete distances. We use an experimentally captured pattern with average occupancy $\sim 1/16$ (as discussed in Sec. ~\ref{sec:theory}) , which is a real-world equivalence of ``$1$ in $4\times 4$'' pattern. We fix the noise settings with $\sigma_r = 0.004$ and $\sigma_s = 0.02$. We set the distance values at $\{8, 10, 12, 14, 16, 18\}$ (the distances are defined such that no image attenuation at distance $=1$), and assume the reference pattern is at distance $= 8$ in calibration. We use a baseline $= 5.0$, which leads to disparities $= \{0, 1.0, 1.7, 2.1, 2.5, 2.8\}$ pixels at the evaluation distances.\footnote{The disparity estimation is not influenced by the reference's distance as long as the disparity is within reasonable range.} 
The evaluation is conducted on a test set with 128 scenes. 
The quantitative comparisons between these two approaches are shown in Fig.~\ref{quantitative} (a, b, c). It can be seen that PF consistently outperforms UF at long distances (green region). When flash power is more attenuated (with a farther imaging distance), the difference between the two approaches is larger. Note that at closer range (e.g., $< 8$),  PF has lower reconstruction quality compared with the UF. We attribute it to the fact that the PF system discards part of the high-frequency information. This is the major limitation of PF technique. For more discussions, please refer to Sec.~\ref{sec:discussion}.

We further evaluate the influence of the baseline. We evaluate PF image reconstruction with a zero baseline (disparity is always zero). As shown by the blue solid lines and dashed lines in Fig.~\ref{quantitative} (a, b, c), the discrepancies in image reconstruction with and without baseline are generally small. When distance increases, performance of the system with baseline drops slightly faster due to more inaccurate depth estimation with lower signal levels. We also evaluate the mean absolute error (MAE) of disparity estimation (Fig.~\ref{quantitative} (d)). The proposed system achieves sub-pixel disparity estimations in the distance range of interest, which provides a rough depth estimation.

\begin{figure}[t!]
\begin{center}
\includegraphics[width=0.9\linewidth]{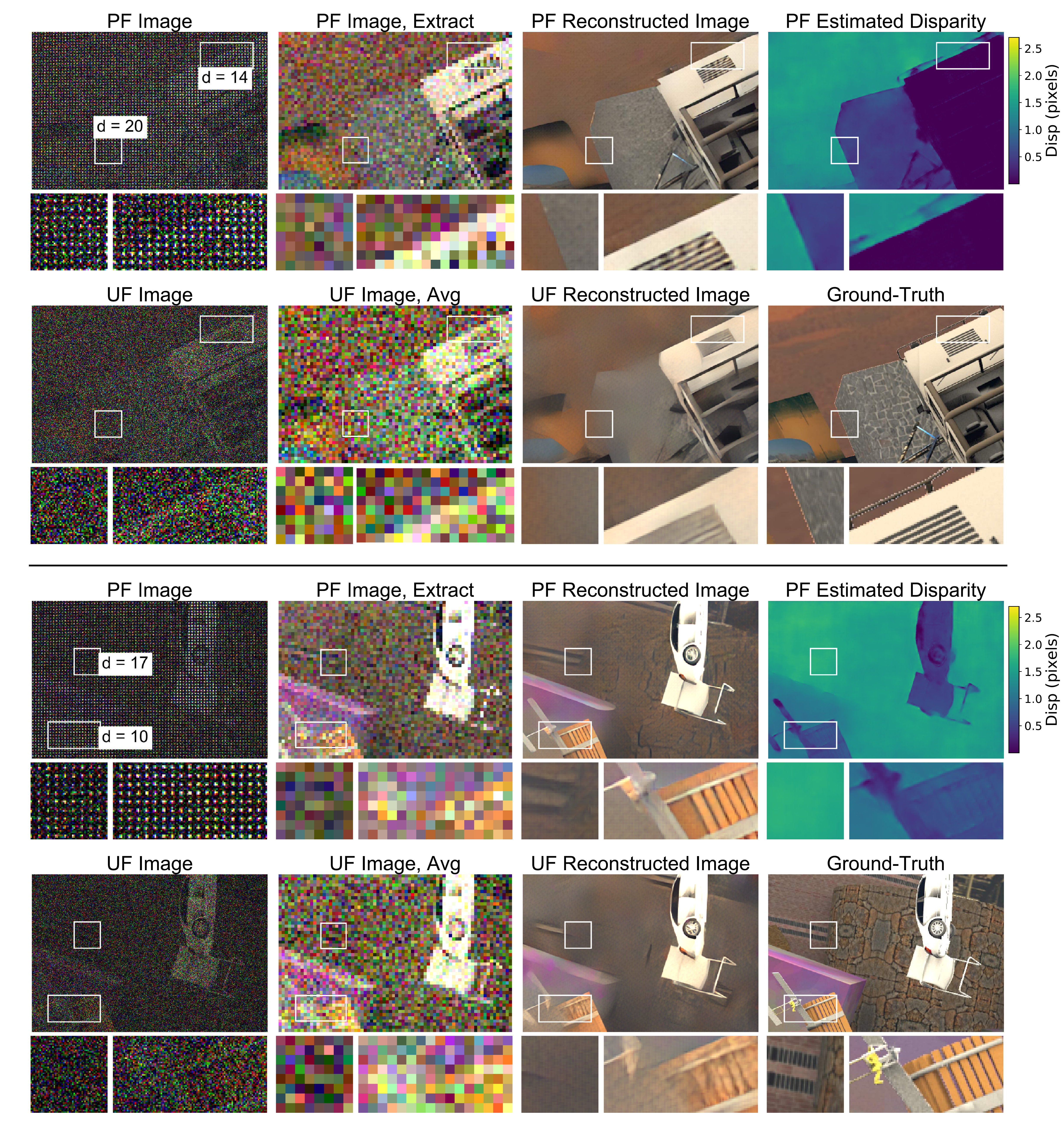}
\end{center}
   \caption{Qualitative comparisons between patterned flash (PF) and uniform flash (UF) on synthetic data. PF preserves more details in the captured image (first and second columns), and supports higher quality reconstruction (third column). PF also provides disparity estimation, which is non-trivial with UF.}
\label{qualitative}
\end{figure}

\begin{figure}[t!]
\begin{center}
\includegraphics[width=0.85\linewidth]{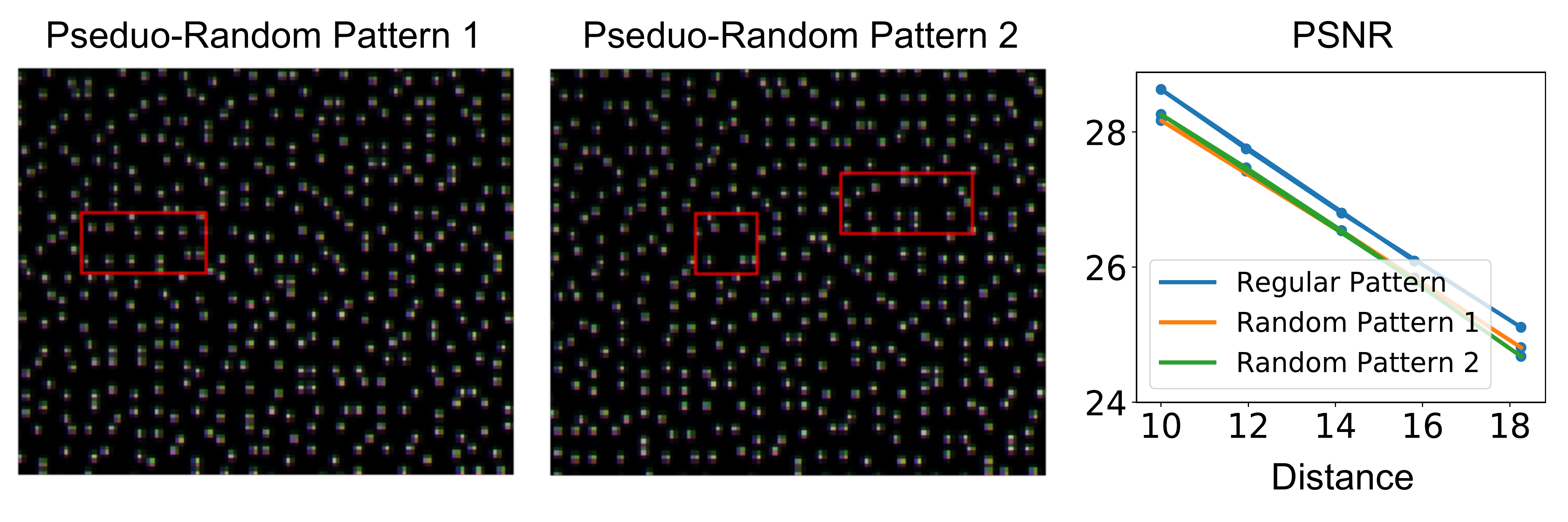}
\end{center}
   \caption{Pattern design:``pseudo-random'' pattern vs. ``regular'' pattern. (a, b) Part of example pseudo-random patterns generated from the regular pattern. Red boxes show big ``gaps'' in the patterns. (c) Quantitative comparison on image reconstruction quality between regular pattern and pseudo-random patterns.}
\label{pattern_design}
\end{figure}

\begin{figure}[t!]
\begin{center}
\includegraphics[width=0.85\linewidth]{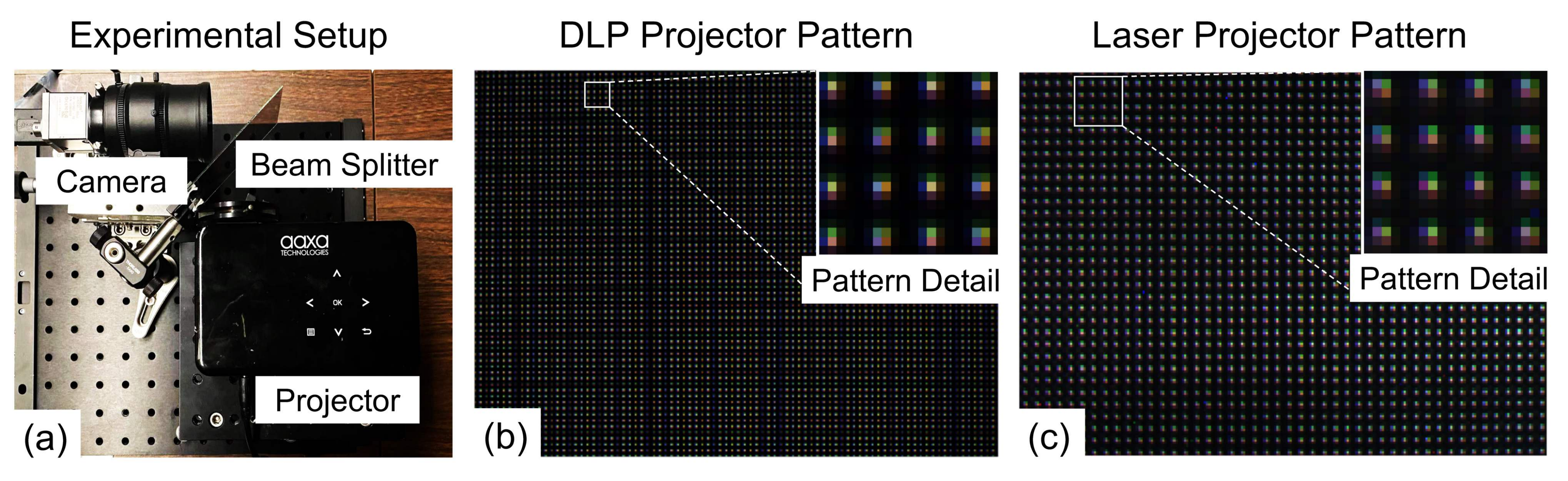}
\end{center}
   \caption{(a) Photo of the hardware prototype, with a camera, a beam splitter (to enable tunable baseline between camera and projector from 0 $\sim$ 2cm), and a projector. (b) DLP projector pattern (with zoomed in details). (c) Laser projector pattern (with zoomed in details).}
\label{setup}
\end{figure}

As discussed in Sec.~\ref{sec:theory}, we can directly compare the \textit{captured} image quality through a ``pixel-binning'' method. The theoretical upper bound of input SNR gain, with the specific pattern being used, is $4\times$ ($12$dB). With the existence of photon shot noise and lower pattern contrast, we get $\sim 6 $dB input SNR gain across the distance range. A qualitative comparison of the ``pixel-binned'' input images is shown in the second column of Fig.~\ref{qualitative}. PF images contain most edges and textures in the scene, while some of these details are invisible in UF images, even after averaging. In this way, PF enables much more detailed reconstructions for far-away scenes (as shown in the third column of Fig.~\ref{qualitative}). We also show the disparity estimation results from PF images (thr fourth column of Fig.~\ref{qualitative}). Note that the algorithm is unable to predict a decent depth map from UF images.

\subsection{Illumination Pattern Design}
Designing optimal patterns in structured light has long been investigated~\cite{optim_si}, \cite{freecam3d}. Since for natural scenes, important details appear with uniform probabilities across the FOV, we also consider patterns with uniformly distributed sampling dots. Current structured light systems utilize random patterns due to less ambiguity in stereo matching. We compare a ``regular'' pattern with multiple ``pseudo-random'' patterns generated from jittering each dot in the regular pattern by $\{-1,0,1\}$ pixels. Two examples of the pseudo-random pattern are shown in Fig.~\ref{pattern_design}(a, b). We empirically found that using pseudo-random pattern results in a slightly worse image reconstruction quality (Fig.~\ref{pattern_design}(c)). This might be due to unavoidable ``gap'' regions in pseudo-random patterns, where information is missing and difficult to be completed (marked by red boxes in Fig.~\ref{pattern_design}(a, b)). Also, in the proposed system and applications, the baseline is much smaller than the scene distance, and disparities are smaller than the unambiguous range in regular patterns. As an example, a $90^{\circ}$ FOV camera, with $2000\times 2000$ pixels, and a baseline $= 3$cm between camera and flash light source, would perceive $2.25$ pixels pattern dot shift when imaging an object at $20$m distance (suppose reference pattern is calibrated at $8$m). 
Therefore, we choose the regular pattern in our setup. More advanced pattern designs (e.g. Poisson disk sampling~\cite{pdisk}) or even an end-to-end optimization on the illumination pattern~\cite{freecam3d,si2} could be left for a future research work.

\begin{figure}[t!]
\begin{center}
\includegraphics[width=0.9\linewidth]{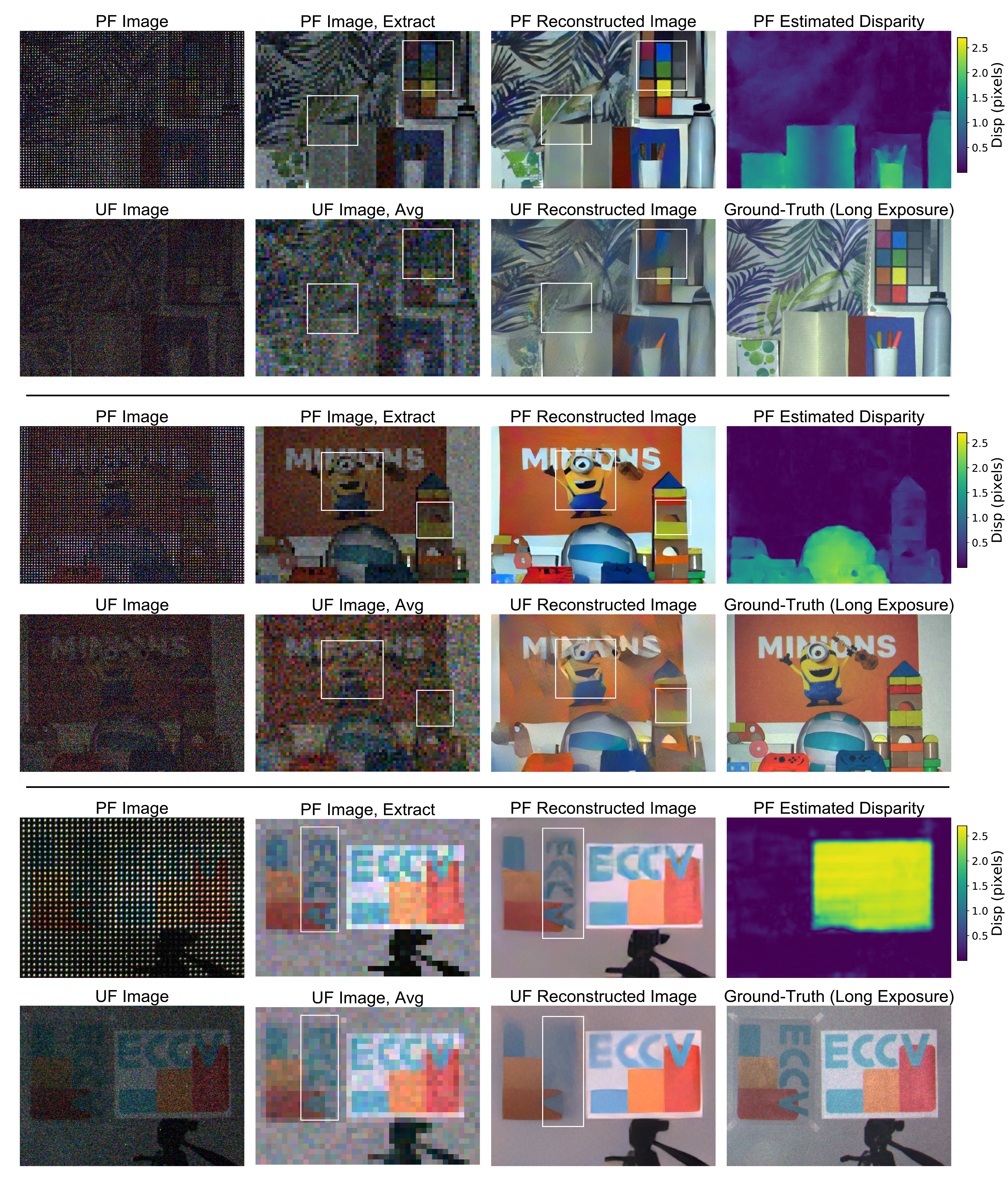}
\end{center}
   \caption{Experimental results of PF vs. UF on real-world data with a DLP projector (first two rows) and a laser projector (third row). Notice that PF resolves more details, especially in the regions indicated by the white boxes.}
\label{real_world}
\end{figure}

\section{Experimental Results}
\label{exp_result}
We implement a patterned flash hardware prototype to evaluate our approach in a real-world environment, as shown in Fig.~\ref{setup}. The system consists of a machine-vision camera (Basler acA2040-120uc), a projector as the PF light source, and a beam splitter. 
The beamsplitter is used to remove the physical constraints from the bulky camera lens and the projector case, and enable tuning the baseline between the camera and the projector from $0$ to $2$cm. We tested our system with two projectors: a DLP projector (Aaxa P300 Pico Projector) and a laser projector (AnyBeam Laser Projector). The DLP projector has a limited depth of focus but with a sharper projected pattern and higher power. The laser projector has a large depth of focus, but has less contrast and lower power.
We set the projector display content as a regular pattern for PF, and as a uniform pattern for the UF. The two patterns are both projected onto a white wall for calibration. We adjust the projector brightness to make the two calibration patterns have the same per-channel average intensities for a fair comparison. We compare both the captured image and the reconstruction qualities in Fig.~\ref{real_world}. In the first setup (Fig.~\ref{real_world} first two rows), the DLP projector projects a pattern with $57\times 80$ dots, and the scene is set to be $\sim 3$m away (limited by lab space), with a depth range $\sim 50$cm (limited by projector depth of focus). In the second setup (Fig.~\ref{real_world}(b)), the laser projector projects a pattern with $28 \times 46$ dots. The scene is set to be $\sim 3$m aways, with depth range $\sim 1.5$m.

\begin{figure}[t!]
\begin{center}
\includegraphics[width=1.0\linewidth]{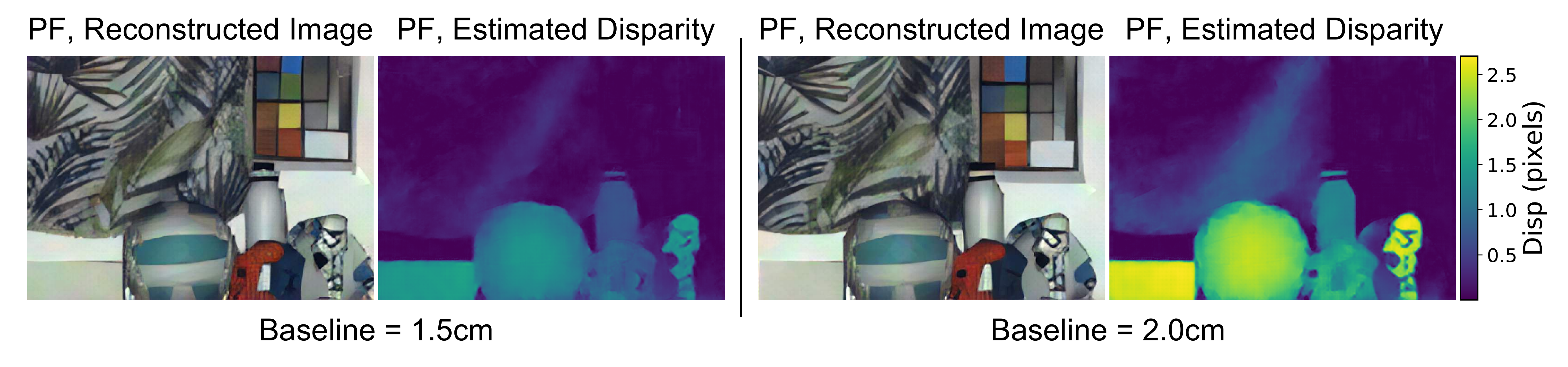}
\end{center}
   \caption{Experimental image reconstruction and disparity estimation with different baselines, with the same camera and DLP projector illumination settings as that in Fig.~\ref{real_world}.}
\label{real_world_depth}
\end{figure}

In Fig.~\ref{real_world}, it can be seen that for both setups, PF out-performs UF, with higher visibility of objects under low flash illumination power, as highlighted by the white boxes. This can be observed in both ``pixel-binned'' images and reconstruction results. Specifically, UF fails to distinguish the outlines of objects (e.g., curtain textures in the first row and toy building blocks in the second row) while PF can faithfully reconstruct the fine textures. In the third row, we placed two boards at different distances. At close range, both PF and UF recovers the board content well enough, due to sufficiently high signal level. At a long distance, UF signals are overwhelmed by the sensor noise, and the reconstructed object is heavily blurred out. The PF system suffers from aliasing artifacts. However, this is a minor effect compared with the artifacts in reconstructed images from the UF. We also demonstrate depth estimation with the experimental setup. As shown in Fig.~\ref{real_world}, we are able to estimate disparities in both regions with rich textures (e.g., curtain, stripes on the baseball) and non-textured regions (e.g., white wall, white notebook). Also, as shown in Fig.~\ref{real_world_depth}, we tune the baseline between the camera and DLP projector to generate different disparities for the same scene. When the baseline changes in a reasonable range, image reconstruction quality is not influenced. 

\begin{figure}[t!]
\begin{center}
\includegraphics[width=1.0\linewidth]{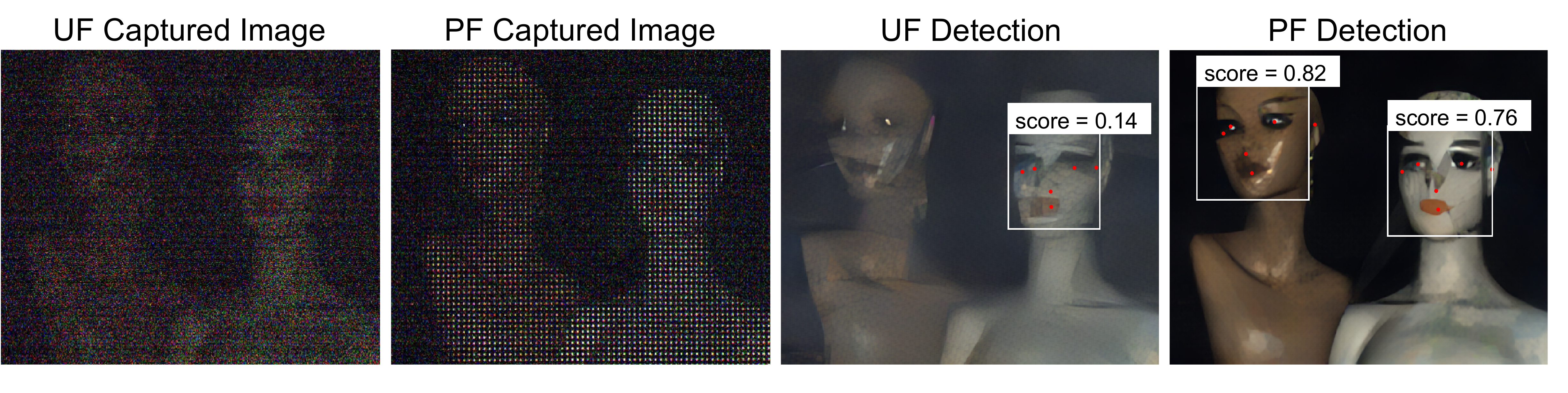}
\end{center}
   \caption{Real-world face detection with UF and PF. The detector fails on UF reconstructed image but succeeds in PF reconstructed image, especially for the mannequin with darker skin color.}
\label{obj_det}
\end{figure}

\textit{Object Detection.}
We further demonstrate the advantage of  PF, by applying face detection on the real-world reconstruction images. We use Google MediaPipe~\cite{mediapipe} as the detector. As shown in Fig.~\ref{obj_det}, the detector is able to both detect the face and the key points from the PF reconstructed image while it fails with the UF reconstructed image, especially for the mannequin with darker skin color. For more discussions and examples of applying PF reconstructed images in downstream tasks, please refer to the supplementary material.

\section{Discussions \& Conclusions}
\label{sec:discussion}
In this paper, we propose patterned flash for low-light imaging at a long distance. From a noisy patterned flash image, we jointly reconstruct a color image and estimate depth. We build a hardware prototype and demonstrate compelling performance under low-light conditions compared to traditional uniform flash. The proposed technique has the potential to be implemented as an energy-efficient low-light imaging module, which improves visual quality and robustness in applications across disciplines, including photography, navigation, and surveillance. 

\textit{Limitations and Envisioning.} (1) In reality, the proposed system can be implemented with more stable and more compact devices. Diffractive optical element (DOE) \cite{NIL}, Vertical-cavity surface-emitting laser (VCSEL) arrays~\cite{Vixar} or Micro-electromechanical systems (MEMS) mirror can be used. (2) As mentioned in Sec.~\ref{sec:simu_result}, at close range, patterned flash has lower image reconstruction quality compared with the uniform flash. This originates from the sparse nature of captured patterned flash signals. We envision two variants of the patterned flash system to solve this problem: (i) a combination of uniform and patterned flashes. Power can be distributed between the two light sources according to the required imaging distance and the relative importance of close or far-away objects. (ii) a lens-based system with a large aperture that focuses at infinity (or the maximum imaging distance of interest), to create patterned flash far away and uniform flash (blurred) nearby. (3) The sparse patterned flash signal might also lead to aliasing in the reconstructed image. However, this is a secondary effect when imaging at long distance, where the uniform flash fails to reconstruct even big structures. Also, when there is certain amount of ambient light or the flash has some uniform flash component, this problem can be relieved. Besides, image inpainting and super-resolution algorithms~\cite{mae}, ~\cite{swinir} can be applied.

\section{Acknowledgement}
We would like to thank Jinhui Xiong and Yifan Peng for helpful discussions.




\clearpage
%
%
\bibliographystyle{splncs04}
\bibliography{eccv2022submission}
\end{document}